# Architectural Pattern of Health Care System Using GSM Networks

[1] Meiappane. A, [2] Dr. V. Prasanna Venkatesan, [3] Selva Murugan. S, [4] Arun. A, [5] Ramachandran. A

*Abstract*—Large-scale networked environments, such as the Internet, possess the characteristics of centralised data, centralised access and centralised control; this gives the user a powerful mechanism for building and integrating large repositories of centralised information from diverse resources set. However, a centralised network system with GSM Networks development for a hospital information systems or a health care information portal is still in its infancy. The shortcomings of the currently available tools have made the use of mobile devices more appealing. In mobile computing, the issues such as low bandwidth, high latency wireless Networks, loss or degradation of wireless connections, and network errors or failures need to be dealt with. Other issues to be addressed include system adaptability, reliability, robustness, extensibility, flexibility, and maintainability. GSM approach has emerged as the most viable approach for development of intelligent software applications for wireless mobile devices in a centralized environment, which gives higher band width of 900 MHz for transmission. The e-healthcare system that we have developed provides support for physicians, nurses, pharmacists and other healthcare professionals, as well as for patients and medical devices used to monitor patients. In this paper, we present the architecture and the demonstration prototype.

## I. INTRODUCTION

Healthcare is a field in which accurate record keeping and communication are critical and yet in which the use of computing and networking technology lags behind other fields.

Healthcare professionals and patients are often uncomfortable with computers, and feel that computers are not central to their healthcare mission, even though they agree that accurate record keeping and communication are essential to good healthcare. In current healthcare, information is conveyed from one healthcare professional to another through paper notes or personal communication. For example, in the United States, electronic communication between physicians and pharmacists is not typically employed but, rather, the physician writes a prescription on paper and gives it to the patient. The patient carries the prescription to the pharmacy, waits in line to give it to a pharmacist, and waits for the pharmacist to fill the prescription. To improve this process, the prescriptions could be communicated electronically from the physician to the pharmacist, and the human computer interfaces for the physicians, nurses, pharmacists and other healthcare professionals could be voice enabled.

## II. E-HEALTHCARE

The e-healthcare system that we have developed is based on the GSM Networks. It provides well-defined interfaces for client applications and separates the interfaces from their implementations. It allows service capabilities and interfaces to be implemented as a collection of processes. GSM Networks provide higher bandwidth of transmission and high clarity to communicate with the physician's PDA.

The Service Functionality of our e-healthcare system consists of three layers as shown in Figure 1. The top layer provides the Web Services interfaces. The bottom layer contains the healthcare services. The Services Coordinator within the middle layer controls the flow of messages in the system from the Web Services interfaces to the healthcare services and vice versa. The GSM architecture which is designed in our approach gives modular support and interoperability over the different Networks and platforms of varying operating system. Personal Health Information is confidential, so access to such information must be restricted to authenticated and authorized users. Secure transmission of such information must be complemented with secure storage of the data [4]. The use of the Web Services is critical for enforcing such policies.

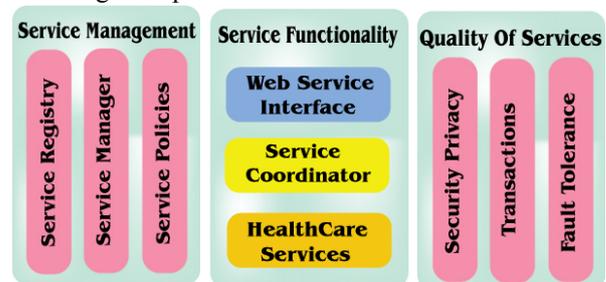

Figure 1. E-Health Care

The security is implemented such that one physician's prescription for a particular patient is not viewed by other physicians but they can view their problems [9]. So this enforces the high level security over the Health Care environment. Also users of the system are authenticated, and session information is kept with logging of service calls.

1 Asst. Professor, 2 Reader, 3, 4,5 Student
1,3, 4,5 Department of Information Technology,
Sri Manakula Vinayagar Engineering College, Puducherry – 605 107. India.
2 Department of Banking Technology, Pondicherry University, Puducherry
1 auromei@yahoo.com, 2 prasanna_v@yahoo.com, 3 selvaa02@rediffmail.com, 4 garunv@gmail.com, 5 itram88@yahoo.com

Resources in the system are attached to the resource creator, and privileged users can view/modify the data in the system.

### III. GSM NETWORKS

A GSM network is made up of multiple components and interfaces that facilitate sending and receiving of signaling and traffic messages. It is a collection of transceivers, controllers, switches, routers, and registers. A Public Land Mobile Network is a network that is owned and operated by one GSM service provider or administration, which includes all of the components and equipment as described below. For example, all of the equipment and network resources that is owned and operated by Cingular are considered a Public Land Mobile Network. Newer versions of the standard were backward-compatible with the original GSM phones.

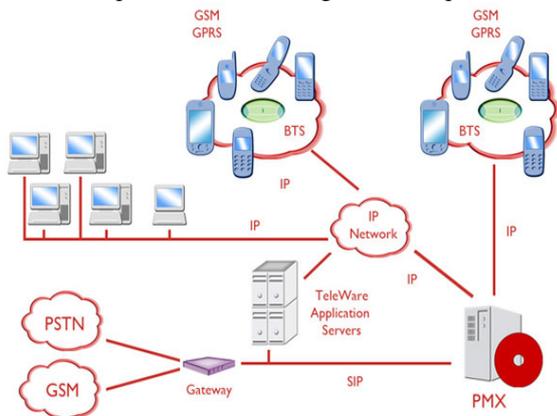

Figure 2.GSM Architecture

GSM is a cellular network, which means that mobile phones connect to it by searching for cells in the immediate vicinity. There are five different cell sizes in a GSM network that is macro, micro, Pico, femto and umbrella cells. The coverage area of each cell varies according to the implementation environment. Macro cells can be regarded as cells where the base station antenna is installed on the clinic infrastructure above average roof top level. Micro cells are cells whose antenna height is under average roof top level; they are typically used in urban area.

Pico cells are small cells whose coverage diameter is a few dozen meters within which the physicians can exchange the patients' details if each one wishes; they are mainly used indoors. Femto cells are cells designed for use in residential or small clinical environments and connect to the service provider's network via a broadband internet connection. Umbrella cells are used to cover shadowed regions of smaller cells and fill in gaps in coverage between those cells present in between two physicians.

The longest distance the GSM specification supports in practical use is 35 kilometers. There are also several implementations of the concept of an extended cell, where the cell radius could be double or even more, depending on the antenna system, the type of terrain and the timing advance. Indoor coverage is also supported by GSM and may be achieved by using an indoor Pico cell base station which is present in the roof top of the clinic, or an indoor repeater with distributed indoor antennas fed through power splitters, to deliver the radio signals containing the information about the patients from an antenna outdoors to the separate indoor distributed antenna system. The modulation used in GSM is Gaussian minimum-shift keying (GMSK), a kind of continuous-phase frequency shift keying. In GMSK, the signal to be modulated onto the carrier is first smoothed with a Gaussian low-pass filter prior to being fed to a frequency modulator.

#### A. GSM Security

GSM was designed with a moderate level of security. The system was designed to authenticate the subscriber using a pre-shared key and challenge-response. The development of UMTS(Universal Mobile Telecommunications System)introduces an optional USIM (UMTS Subscriber Identity Module) that uses a longer authentication key to give greater security, as well as mutually authenticating the health care network and the user - whereas GSM only authenticates the physicians to the network (and not vice versa). The security model therefore offers confidentiality and authentication, but limited authorization capabilities, and no non-repudiation. GSM uses several cryptographic algorithms for security.

The A5/1 and A5/2 stream ciphers are used for ensuring over-the-air voice privacy. A5/1 was developed first and is a stronger algorithm used within Europe and the United States; A5/2 is weaker and used in other countries. Serious weaknesses have been found in both algorithms: it is possible to break A5/2 in real-time with a cipher text-only attack, and in February 2008, Pico Computing, Inc revealed its ability and plans to commercialize FPGAs that allow A5/1 to be broken with a rainbow table attack. The system supports multiple algorithms so operators may replace that cipher with a stronger one.

#### B. GSM Frequencies

GSM networks operate in a number of different frequency ranges. Most 2G GSM networks operate in the 900 MHz or 1800 MHz bands. The PDA which physicians handle can access the data from the server in a less time so that our system is said to have a high latency. Most 3G GSM networks operate in the 2100 MHz frequency band. The rarer 400 and 450 MHz frequency bands are assigned in some countries where these frequencies were previously used for first-generation systems. GSM-900 uses 890–915 MHz to send information from the mobile station that is from the physician to the base station (uplink) and 935–960 MHz for the other direction that is the patients or to the pharmacist (downlink), providing 125 RF channels (channel numbers 1 to 124) spaced at 200 kHz[7]. Duplex spacing of 45 MHz is used.

Time division multiplexing is used to allow eight full-rate or sixteen half-rate speech channels per radio frequency channel to send the data about the patients. There are eight radio timeslots (giving eight burst periods) grouped into what is called a TDMA frame. The channel data rate for all 8 channels is 270.833 kbit/s, and the frame duration is 4.615 ms. The transmission power in the handset of the physician i.e. PDA is limited to a maximum of 2 watts in GSM850/900 and 1 watt in GSM1800/1900. GSM uses several cryptographic algorithms for security.

*C. The Advantages of 900-MHz Bandwidth*

The AT&T has offered 3G services in the 850-MHz frequency for years; this has been available in major cities. The major advantage of this 900 MHz is that it provides better power consumption compared to other 3G technologies available. However, the vast majority of users have already migrated to 3G networks. As well as allowing cheaper and relatively easy upgrades of existing GSM network infrastructure to provide 3G mobile broadband services, the licensing terms of the 900MHz bandwidth state that allocation won't be dependent on a particular format or technology—meaning the spectrum can be used to provide upcoming 4G services like long-term evolution (LTE).

IV. HEALTHCARE SERVICES

The healthcare services of our e-healthcare system are provided by Clinic architecture and Pharmacy architecture. The system also provides user interfaces for patients, physicians, nurses, and pharmacists. The devices accessing these architectures can be desktop or server computers and PDAs or smart cell phones having GSM Networks which improve higher bandwidth of transmission from PDA to the patient.

*A. Architectural pattern of the Clinic:*

The Clinic Architecture exposes two interfaces, a Web Server and a Web Service which is connected by the GSM Networks, for the clinic staff, the patients and the medical monitoring devices, as shown in Figure 4.

The Web Server interface is intended for users who prefer to use a Web browser to access the healthcare services via GSM Networks. The Web Server uses the GSM Networks to access the data which enables a higher bandwidth. The Clinic architecture provides support for routine activities of the physician. It maintains information, such as the physician's appointments for a specific day/week as shown in Figure 3, the patients that he/she has examined, notes related to the patients, etc. Access to a patient's private information is restricted and secured, as discussed previously.

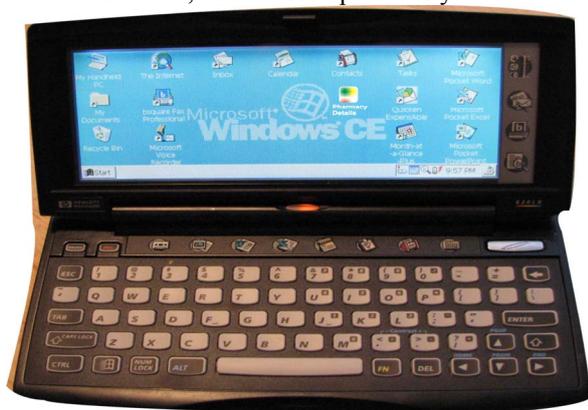

Figure 3. PDA used by the Physicians.

The Clinic architecture sends prescriptions from the physician to the desired pharmacies over the GSM Networks using the Web Service provided. It uses the Yahoo! Local Search Web Service to locate the pharmacy closest to the patient's home or the physician's office.

The physician can use the Web Server interface to access the e-healthcare system using a browser from a desktop computer or a PDA via GSM Networks. The physician can use the PDA to enter/retrieve information about the patient during/after an appointment and access this information any time, any where. The use of a PDA with a small keyboard makes it difficult for the physician to input information about the patient.

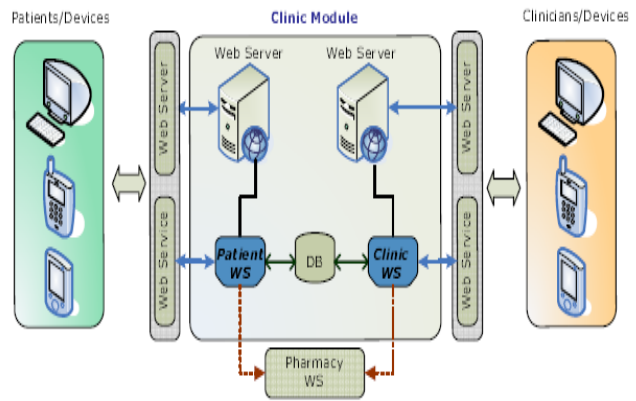

Figure 4. Clinic Architecture.

*B. Architectural pattern of the Pharmacy:*

The Pharmacy Architecture exposes Web Server and Web Service interfaces. The Web Server interface allows the users to access the e-healthcare system at the pharmacy using a browser. The Web Service interface provides access for applications deployed at the pharmacy and can also be used by humans and devices.

The Pharmacy architecture provides services to the pharmacist, patients and devices used at the pharmacy. The Pharmacy Architecture keeps a record of the patient's prescriptions for the pharmacist's and the patient's reference. When the physician submits a new prescription to the pharmacy, the Clinic architecture at the physician's office communicates with the Pharmacy architecture at the pharmacy over the GSM Networks. Removing human intervention from the communication between the physician and pharmacist, and maintaining the information electronically, reduces the possibility of human errors.

The pharmacist can view the outstanding prescriptions for the patients, as they are received from the physicians. The Pharmacy architecture updates the status of the prescriptions as the pharmacist fills them. The patient can determine, via the GSM Networks, whether a prescription has been filled and is ready for pick up or delivery. Most of these prescriptions are renewals of existing prescriptions via the PDA and GSM Networks. Therefore, the patient interface also has access to services that provide renewal of existing prescriptions, custom alerts, etc.

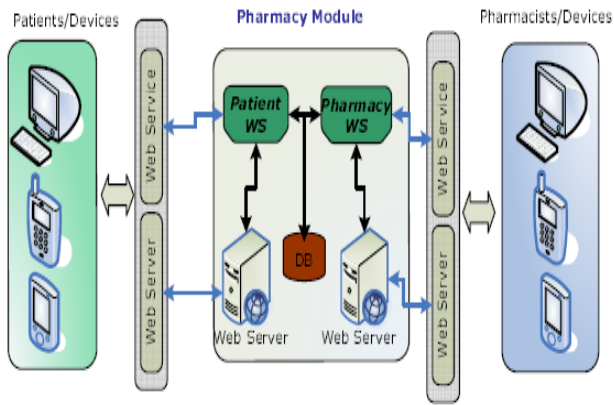

Figure 5. Pharmacy Architecture.

## V. ARCHITECTURE – A MULTI SERVER

Managing clinical information is a challenge with unique requirements, and so far, no system has been able to address the complexity of the entire hospital environment. Some medical information systems such as Hospital Information System (HIS), Picture Archiving and Communication System (PACS), Radiology Information System (RIS) and Laboratory Information System (LIS) are used in hospitals now, but they are usually heterogeneous and isolated. Data is incomplete, workflow is discontinuous, and management is not uniform.

Therefore, it has been medical staffs' dream to establish the Enterprise Hospital Information System that could integrate all heterogeneous systems and make all clinical data including clinical report, lab results and medical images available whenever and wherever they are needed. Some of them define standardized interfaces to many healthcare "Object Oriented Services" such as CORBAmed (Common Object Request Broker Architecture in Medicine), which realizes the share of common functionalities like access control among different systems.

The following properties of distributed systems make them increasingly essential as the foundation of information and control systems:

*a) Collaboration and connectivity:* An important motivation for distributed systems is their ability to connect us to vast quantities of geographically distributed information about the patients and their health information, such as their photo copy of the scanned part and their prescriptions. The popularity of instant messaging and chat rooms on the Internet highlights another motivation for distributed systems: keeping in touch with physicians, patients and the pharmacists.

*b) Economics:* Computer networks that incorporate PDAs, laptops, PCs, and servers often offer a better price/performance ratio than centralized mainframe computers. For example, they support decentralized and modular applications that can share expensive peripherals, such as high-capacity file servers which contain the databases like the HIS, LIS, PACS, CIS and higher solution printers which prints the X-ray of the patients.

Similarly, selected application components and services can be delegated to run on nodes with specialized processing attributes, such as high-performance disk controllers, large amounts of memory, or enhanced floating-point performance which evens the low level health care system can afford.

*c) Performance and scalability:* Successful software typically collects more users and requirements over time, so it is essential that the performance of distributed systems can scale up to handle the increased load and capabilities on the various servers like the HIS, LIS, PACS, CIS. Significant performance increases can be gained by using the combined computing power of networked computing nodes.

In addition—at least in theory— multiprocessors and networks can scale easily. For example, multiple computation and communication service processing tasks can be run in parallel on different nodes in a server farm or in different virtual machines on the same server.

*d) Failure tolerance:* A key goal of distributed computing is to tolerate partial system failures. For example, although all the nodes in a clinic may be live, the network itself may fail. Similarly, an end-system in a network, or a CPU in a multiprocessor system, may crash.

Such failures should be handled gracefully without affecting all—or unrelated—parts of the system. A common way to implement fault tolerance is to replicate services across multiple nodes and/or networks.

*e) Inherent distribution:* Some applications are inherently distributed, including telecommunication management network (TMN) systems, enterprise business systems that span multiple company divisions in different regions of the world, peer-to-peer (P2P) content sharing systems, and business-to-business (B2B) supply chain management systems. So, in that way our clinical module system also can be an inherent distributed system. Distribution is not optional in these types of systems—it is essential to meet customer needs (patients needs).

Stating the above properties of the distributed system, however we see some problems like dealing the networking issues directly and support a location-independent interaction between the physicians. So, Introduce a BROKER to allow distributed systems to find, access, and communicate with one another in the same way as if both parties were collocated. Local brokers on each network node on the hospital system negotiate and perform all interprocess communication on behalf of the system's database.

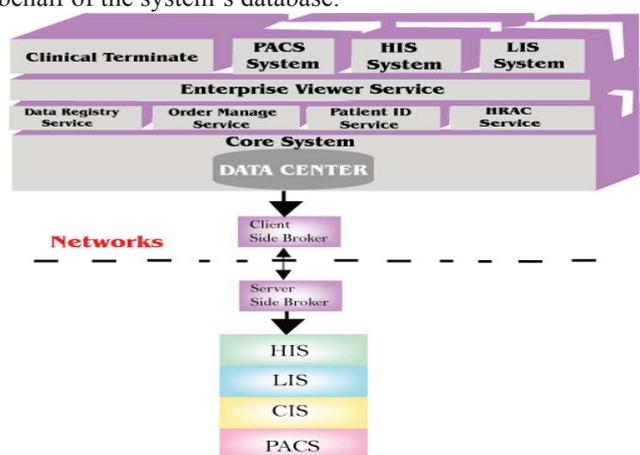

Figure 6. The Enterprise Digital Hospital System

The two key advantages of BROKER architecture are encapsulation and location independence. Encapsulation enables physicians to focus on providing useful clinical functionality: they do not need to bother with low-level networking issues. Location independence allows patients to access remote physicians in the same manner as clinical objects collocated and also in vice-versa inside a computer network.

Location independence also has a positive impact on the system's scalability and availably, because it can take advantage of the collective computing power that is available in the network, for example by means of replication and federation of clinical objects.

All three properties of location independence are especially important in the context of physician-patients architectures, because different physician-patients instances have different concrete functional and operational requirements. The broker we used here are the client side broker and the server side broker which helps to access the data from the multi server in a time less than the usual sever access time, because of this the latency of the system is improved. So, this ease the work done by the physician by the help of the PDA connected throughout the clinical architecture via the GSM networks.

This article proposes an architecture design that deploys GSM Networks to accomplish the specified task. This design achieves the goal of complete integration by embedding all the servers like HIS DB, CIS DB, LIS DB, PACS DB into a single enterprise server so that the access time and the cost per server for the enterprise is reduced.

## VI. THE DESIGN OF ARCHITECTURE

In order to establish an enterprise hospital information system, the integration among these heterogeneous systems must be considered. Complete integration should cover three aspects: data integration, function integration and workflow integration. However most of the previous design of architecture did not accomplish such a complete integration[1]. This architecture design of the enterprise hospital information system based on the concept of GSM deployment in the network.

## VII. IMPLEMENTATION

In our implementation we use the GSM concept and deployed the Web Services on 3 GHz computers with 2 GB memory. We deployed the patient applications on 2 GHz computers with 2 GB memory plus GSM Networks. The patient applications communicate with the Web Server or Web Services over the GSM Networks. For the PDA we used the OQO device, which is a full-featured 3" x 5" personal computer connected via GSM Networks, which is powerful enough to run both an embedded speech recognition engine and our physician application in GSM Networks.

The device features an 800 x 480 resolution screen that is capable of providing the user with detailed graphical information. The PDA can communicate with a desktop or server computer using GSM Networks.

### A. Speech Software

In our e-healthcare system we use SRI's DynaSpeak speech recognition engine. DynaSpeak supports multiple languages, adapts to different accents, and does not require training prior to use. It incorporates a Hidden Markov Model (HMM) for separating speech from interfering signals with different statistical characteristics. DynaSpeak is ideal for embedded platforms, because of its small footprint (Less than 2 MB of memory) and its low computing Requirements (66 MHz Intel x86 or 200 MHz Strong Arm processor).

DynaSpeak can be used with either a finite state grammar or a free-form grammar, which enables patients to use this software even though he/she is an illiterate.

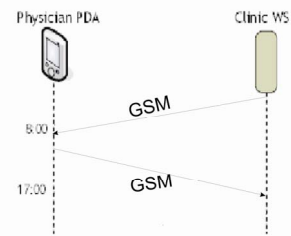

Figure 7. Synchronization of Information on hysician PDA and Desktop/Server Computer

### B. Atom/RSS

It is possible that at certain times or places no network communication is available between physicians, nurses and pharmacists. For example, in an emergency, such as an earthquake, a physician cannot expect to have a network connection to his/her desktop or server computer. Physicians, nurses and pharmacists must be able to access, and modify, healthcare information offline.

Atom/RSS are syndication technologies, based on XML, that enable the sharing and communication of information between heterogeneous platforms by making the information self-describing. We have developed a Consistent Data Replication (CDR) and Reliable Data Distribution (RDD) infrastructure that replicates information from one computer to another using Atom/RSS feeds. We can use this infrastructure to synchronize information on the physician's desktop or server computer with that on his/her PDA, as shown in Figure 7, allowing the physician to view that information when it is offline. At the start of the day, our software on the PDA retrieves the necessary updates from the Clinic Web Service on the desktop or server computer via GSM Networks. At the end of the day, our software on the PDA Generate an update feed for the Clinic Web Service on the desktop or server computer to read.

## VIII. PERFORMANCE EVALUATION

For the performance evaluation we measured the latency of the Pharmacy Web Service, *i.e.*, the delay between the physician's sending a prescription to the pharmacist and the physician application receiving an acknowledgement that the pharmacy application received the prescription. In the experiments we used different rates of sending prescriptions and different numbers of medicines in a prescription. Figure 8 shows the experimental results, namely, the mean latency for a prescription containing a certain number of medicines, taken over 300 seconds. As expected, increasing the rate of sending prescriptions increases the latency. The increased latency is due to the increased usage of the Pharmacy Web Service.

Likewise, for a particular rate of sending prescriptions, increasing the number of medicines in a prescription increases the latency, because the message size is larger and the processing time for the XML is greater.

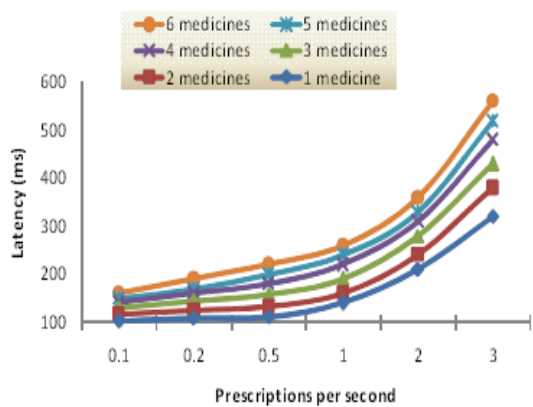

Figure 8. Latency of Pharmacy Web Service.

## IX. RELATED WORK

Extensive work has been undertaken on the development of e-healthcare systems. As key issues, they identify flexibility, adaptability, robustness, integration of existing systems and standards, semantic compatibility, security and process orientation. They define workflow properties for common healthcare practices and present a summary of workflow properties and requirements.

Much of the work on e-healthcare systems has focused on record keeping and databases Work has also focused on access and security as well as on social implications of recording and communicating healthcare information. Less work has been done on human computer interfaces and usability by healthcare professionals and patients. Our e-healthcare system aims to reduce human errors by exploiting electronic communication and record keeping, and by providing user friendly input and output capabilities along with authentication security.

## REFERENCES

[1] Meiappane.A, Maheswaran.S, Prabhakaran.M, Lakshmi Narayanan.A, "Mobile Agent Architecture for Networking in Hospital Organizations and Healthcare Enterprises "International Journal of Computer Theory and Engineering, Vol. 1, No.3 August 2009, 209-215.
[2] W. M. Omar and A. Taleb-Bendiab, "Service oriented architecture for e-health support services based on grid computing," Proceedings of the IEEE International Conference on Services Oriented Computing, Chicago, IL, September 2006, 135-142.
[3] D. Budgen, M. Rigby, P. Brereton and M. Turner, "A data integration broker for healthcare systems," IEEE Computer 40, 4, April 2007, 34-41.
[4] M. Subramanian, A.S. Ali, O. Rana, A. Hardisty and E. C. Conley, "Healthcare@Home: Research models for patientcentered healthcare services," Proceedings of the 2006 International Symposium on Modern Computing, October 2006, 107-113.
[5] Elad Barkan, Eli Biham, Nathan Keller, " Instant Ciphertext only Cryptoanalysis of GSM Encrypted Communication".Proceedings of Crypto May, 2003.http://cryptome.org/gsm-crack-bbk.pdf.
[6] Theodore S. Rappaport "Wireless Communications Principles and Practice", Second edition, Prentice Hall India, 2004.
[7] P. Stuckmann, "The GSM Evolution" John Wiley & Sons, Hoboken, NJ, USA, 2003.
[8] V. Alexandria, "Number of independent pharmacies on the rise, Impact of Medicare Part D looms on The horizon,"April28,2006 http://www.ncpanet.org/media/releases/2006/number of independent pharmacies on the 04-28-2006.php
[9] M. Subramanian, A.S. Ali, O. Rana, A. Hardisty and E. C.Conley, "Healthcare@Home: Research models for patientcentered healthcare services," Proceedings of the 2006 International Symposium on Modern Computing, October 2006, 107-113.
[10] Lu Xudong, Duan Huilong, "A research on healthcare integrating model of medical information system," Journal of Biomedical Engineering China, Vol.22, pp. 108-112, January 2005.
[11] Communications Direct,"EV approves Release of 900Mhz Spectrum for Mobile Broadband" https://www.communicationsdirectnews.com/do.php/130/36449.
[12] M. Beyer, K. A. Kuhn, C. Meiler, S. Jablonski and R. Lenz, "Towards a flexible, process-oriented IT architecture for an integrated healthcare network," Proceedings of the 2004 ACM Symposium on Applied Computing, Nicosia, Cyprus, March 2004, 264-271.

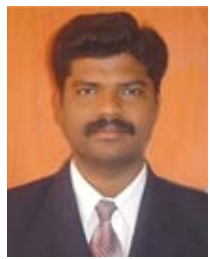

**A. Meiappane** received his M.Tech. degree in Computer Science & Engineering from Pondicherry University, Puducherry, India. He is Currently working as Assistant Professor, Dept. of IT at SMVEC (Pondicherry University), Puducherry. His research interest are in the areas of Wireless and Wireline Networks, Communication Networks and Distributed Systems, Software Engineering & Metrics.
Mr. A. Meiappane is a member of IACSIT. He published in various reputed National and International Conferences and Journals.

**Dr. V. Prasanna Venkadesan** received his M.Tech. and Ph.D degree in Computer Science & Engineering from Pondicherry University, Puducherry, India. He is Currently working as Reader, Dept. of Banking Technology at Pondicherry University, Puducherry. His research interest are in the areas of Software Architecture, Object-Oriented Systems, Multilingual-based systems, Programming Languages, Banking Technology Mgt.

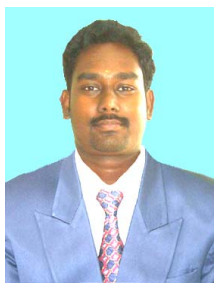

**S. Selva Murugan** is pursuing his B.Tech. degree in Information Technology, Sri Manakula Vinayagar Engineering College, Pondicherry University, Puducherry, India. His research area includes Database Systems and Distributed Systems
S.Selva Murugan presented a paper in ICMLC 2010.

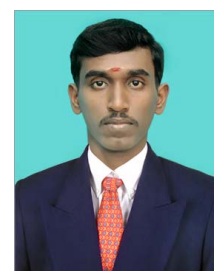

**A. Arun** is pursuing his B.Tech degree in Information Technology, Sri Manakula Vinayagar Engineering College, Pondicherry University, Puducherry, India. His research area includes Networks and Database Systems
A. Arun published a paper in ICMLC 2010.

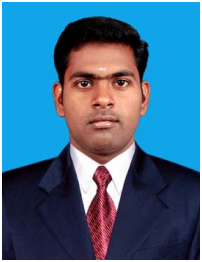
**A. Ramachandran** is pursuing his B.Tech degree in Information Technology, Sri Manakula Vinayagar Engineering College, Pondicherry University, Puducherry, India. His research area includes Networks and Software Engineering
A. Ramachandran published a paper in ICMLC 2010.